\documentclass[prl,superscriptaddress,twocolumn,aps,showpacs]{revtex4}
\usepackage{epsfig}% Include figure files
\usepackage{dcolumn}% Align table columns on decimal point
\usepackage{bm}% bold math

\begin{document}

\title{Ising Quantum Hall Ferromagnet in Magnetically Doped Quantum Wells}

\author{J. Jaroszy\'nski}\email{jaroszy@magnet.fsu.edu}
\affiliation{Institute of Physics, Polish Academy of Sciences, al.\
Lotnik\'ow 32/46, 02--668 Warszawa, Poland} \affiliation{National
High Magnetic Field Laboratory, Florida State University,
Tallahassee, Florida 32310}
\author{T.\ Andrearczyk}
\affiliation{Institute of Physics, Polish Academy of Sciences, al.\
Lotnik\'ow 32/46, 02--668 Warszawa, Poland} \affiliation{National
High Magnetic Field Laboratory, Florida State University,
Tallahassee, Florida 32310}
\author{G.\ Karczewski}
\affiliation{Institute of Physics, Polish Academy of Sciences, al.\
Lotnik\'ow 32/46, 02--668 Warszawa, Poland}
\author{J.\ Wr\'obel}
\affiliation{Institute of Physics, Polish Academy of Sciences, al.\
Lotnik\'ow 32/46, 02--668 Warszawa, Poland}
\author{T.\ Wojtowicz}
\affiliation{Institute of Physics, Polish Academy of Sciences, al.\
Lotnik\'ow 32/46, 02--668 Warszawa, Poland}
\author{E.\ Papis}
\affiliation{Institute of Electron Technology, al. Lotnik\'ow 32/46,
02--668 Warszawa, Poland}
\author{E. Kami\'nska}
\affiliation{Institute of Electron Technology, al. Lotnik\'ow 32/46,
02--668 Warszawa, Poland}
\author{A.\ Piotrowska}
\affiliation{Institute of Electron Technology, al. Lotnik\'ow 32/46,
02--668 Warszawa, Poland}
\author{Dragana Popovi\'c}
\affiliation{National High Magnetic Field Laboratory, Florida State
University, Tallahassee, Florida 32310}
\author{T.\ Dietl}
\affiliation{Institute of Physics, Polish Academy of Sciences, al.\
Lotnik\'ow 32/46, 02--668 Warszawa, Poland}
\begin{abstract}
We report on the observation of the Ising quantum Hall ferromagnet
with Curie temperature $T_C$ as high as 2~K in a modulation-doped
(Cd,Mn)Te heterostructure. In this system field-induced crossing of
Landau levels occurs due to the giant spin-splitting effect.
Magnetoresistance data, collected over a wide range of temperatures,
magnetic fields, tilt angles, and electron densities,  are discussed
taking into account both Coulomb electron-electron interactions and
s$-$d coupling to Mn spin fluctuations. The critical behavior of the
resistance ``spikes'' at $T \rightarrow T_C$ corroborates theoretical
suggestions that the ferromagnet is destroyed by domain excitations.
\end{abstract}
%PACS:73.43.Nq(QHE:quantum phase transitions),72.25.Dc(spin-polarized
%transport in semi), 73.61.Ga,(II-VI)75.50.Pp(Magnetic semi)
 \maketitle

Over the last several decades two-dimensional electron systems
(2DES) have emerged as a principal medium for studying  phenomena
driven by many-body correlation effects since the carrier density
and the relative strength of orbital and spin effects can be
controlled externally. Owing to the quenching of the kinetic
energy, correlation effects are particularly strong in the magnetic
field leading to an unexpectedly rich variety of ground states and
quasiparticle forms \cite{Dass97}. In particular, if Landau levels
(LLs) corresponding to the opposite spin orientations of
quasi-particles at the Fermi level overlap, the spin degree of
freedom is {\em not} frozen by the field so that a spontaneous spin
order may appear at low temperatures \cite{Girv00}, the resulting
state being known as the quantum Hall ferromagnet (QHFM).
Importantly, the ground state is predicted to have the uniaxial
anisotropy if the spin subbands involved originate from different
LLs \cite{Jung00}. The level arrangement corresponding to such an
Ising QHFM has been realized in various III-V 2DES
\cite{Piaz99,Poor00}.

In this Letter we present results of experimental studies, which
demonstrate the existence of a QHFM in a quantum well (QW) containing
magnetic ions. In our modulation-doped n-type (Cd,Mn)Te structures,
owing to the mean-field part of the s$-$d exchange interaction
between the electrons and Mn spins, the spin-splitting of electronic
states is not only giant but depends, in a nonlinear fashion, on the
magnetic field $B$. Accordingly, many crossings of Landau spin
sublevels occur, even without tilting the field direction. The
ferromagnetic ordering gives rise to the presence of hysteresis and
resistance "spikes", making it possible to determine the phase
diagram of a QHFM as a function of the carrier density and tilt
angle. Critical behavior of the spike resistance is found, verifying
the recent theoretical prediction \cite{Jung01}. At the same time,
the Curie temperature $T_C$ is shown to reach 2~K, a value much
higher than that observed and explained theoretically in the case of
both (i) high electron mobility AlAs QW in the quantum Hall effect
(QHE) regime ($T_C \lesssim 0.5$~K \cite{Poor00,Jung01}) and (ii)
n-type diluted magnetic semiconductor (DMS) at $B =0$ ($T_C\lesssim
0.2$~K in (Zn,Mn)O:Al epilayers \cite{Andr01}). This enhanced
stability of the QHFM phase is rather surprising in view of both
previous results \cite{Poor00,Jung01} and the significance of
disorder in the present material. It is discussed in terms of
electron-electron interactions \cite{Jung00} as well as by
considering the role of static and dynamic fluctuations in the
subsystem of the Mn spins.

Previous magnetotransport studies of 2DES containing a DMS channel
have demonstrated that the Mn spins have no effect on the precision
of the Hall resistance quantization \cite{Grab93,Smor97,Jaro00}.
Furthermore, the high degree of spin polarization has made it
possible to verify the temperature and size QHE scaling up to the
filling factor as high as $\nu=8.5$ \cite{Jaro00}, avoiding
uncertainties associated with the overlap of spin-split LLs. However,
some anomalies of the QHE, which were absent in CdTe QW, have been
noted in these studies \cite{Jaro00}. In view of the present results,
they were  assigned correctly to crossings of LLs.

The modulation-doped (Cd,Mn)Te QW was grown for the present study
by molecular beam epitaxy, exploiting the previous expertise
\cite{Karc98,Jaro00} on how to obtain 2DES with an adequate carrier
density and mobility in a QW with a sizable effective Mn
concentration.  Barriers of Cd$_{0.8}$Mg$_{0.2}$Te were separated
from (001) undoped GaAs substrate by 1~nm ZnTe and 3~$\mu$m CdTe
buffer layers. The Mn ions were inserted into the QW by the digital
alloy technique, that is by depositing three evenly spaced
monolayers of Cd$_{1-x}$Mn$_x$Te during the growth of CdTe QW with
the thickness $L_W=10$~nm, as depicted schematically in the inset
to Fig.~1(a). According to the spectral position of the
photoluminescence line at 77~K, average $x=2 \pm 0.5$\%. The iodine
donors were introduced to the front barrier at 20~nm away from the
QW. For magnetotransport measurements a $0.5\times 1$~mm$^2$ Hall
bar was etched, to which leads were soldered with indium. A gold on
chromium front gate was deposited onto cap layer, and makes it
possible to control the electron density $n_s$ between 1.6 and
$3.6\times10^{11}$~cm$^{-2}$, as determined by Hall measurements at
10 and 23 K. The low-temperature electron mobility, limited to a
large extent by short-range alloy and spin disorder scattering, is
about $\mu = 1.6\times10^4$ cm$^2$/Vs. The experiments were
performed in magnetic fields up to 18~T and down to temperature $T$
of either 0.24~K in a pumped $^3$He system or 0.03~K in a
$^3$He/$^4$He dilution refrigerator. In the latter, the sample can
be rotated in order to change the angle $\theta$ between the
interface normal and ${\bm B}$.

%%%%%%%%%%%%%%% figure 1 %%%%%%%%%%%%%%%%%%%%%%%%%%%%%%%%%%%%%%%%%%%%
\begin{figure}[t]

\centerline{\epsfig{file=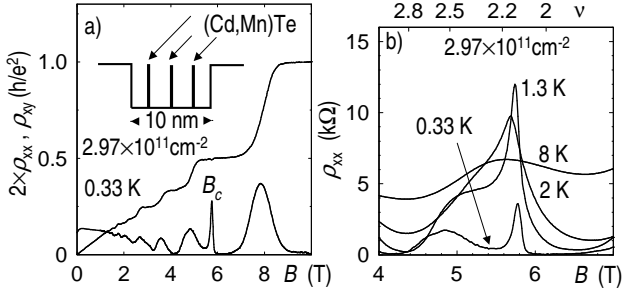,width=8.2cm,clip==}} \caption{
(a) Resistances $\rho_{xx}$ and  $\rho_{xy}$ at $T=0.33$~K for
$n_s=2.97\times10^{11}$~cm$^{-2} (V_g =0)$. Note the presence of a
spike in $\rho_{xx}$ at $B_c\simeq 5.8$~T, shown at selected
temperatures in (b). Inset: the nominal distribution of magnetic
monolayers in the QW.}
\end{figure}
%%%%%%%%%%%%%%%%%% end figure 1%%%%%%%%%%%%%%%%%%%%%%%%%%%%%%%%%%%%%%

Figure 1(a) presents the Hall $\rho_{xy}$ and the longitudinal
$\rho_{xx}$ resistivities,  the latter revealing the presence of a
strong resistance "spike" in the magnetic field $B_c\approx 5.8
$~T, which corresponds to the LL filling factor $\nu\equiv n_s h/eB
\approx 2$. According to Fig.~1(b) the $\rho_{xx}$ anomalies peak
around 1.3~K. Moreover, at the same $T$ a dramatic shift of the
Shubnikov-de Haas (SdH) maxima is clearly visible in Fig.~1(b).
Figure~2(a) shows, in turn, that $B_c\cos\theta$ decreases  when
$\theta$ increases. Guided by the previous experimental
\cite{Poor00} and theoretical \cite{Jung01} results, we presume
that the spikes appear at the crossing points of spin sublevels. In
order to evaluate the field values $B_c$ at which crossings are
expected in our system, we start from the known form of the LL
energy in DMS \cite{Diet94},
\begin{eqnarray}\label{Brillouin}
E_{n,\uparrow,\downarrow}=(n+1/2)\hbar eB\cos\theta/m^* \pm
\nonumber
\\ \pm\frac{1}{2}\left[g^*\mu_BB + \alpha
N_0x_{eff} S{\textrm
B}_{S}\left(\frac{g\mu_BB}{k_B[T+T_{AF}]}\right)\right].
\end{eqnarray}
Here $n$ is the LL index; $m^* = 0.10m_0$ and $g^*=-1.67$
\cite{Hu98} are the effective mass and Land\'e factor of the
electrons in CdTe; $\alpha N_0=0.22$~eV is the s$-$d exchange
energy \cite{Diet94,Gaj94}, and B$_S$ is the Brillouin function, in
which $S=5/2$ and $g = 2.0$. The functions $x_{eff}(x) < x$ and
$T_{AF}(x)>0$, known from extensive magnetooptical studies of
uniform Cd$_{1-x}$Mn$_x$Te layers \cite{Gaj94}, describe the
reduction of magnetization $M(T,H) = g\mu_Bx_{eff}N_0S{\textrm
B}_{S}(T,H)$ by antiferromagnetic interactions.

The crossing occurs at $E_{1,\downarrow}-
E_{0,\uparrow}=\varepsilon(B\cos\theta)$, where $\varepsilon$ is an
energy correction brought about by an electron exchange interaction
with the occupied $0\downarrow$ LL. We take $\varepsilon$ in the
form determined experimentally \cite{Poor00} and theoretically
\cite{Jung01} for the AlAs QW, $\varepsilon$[K] $=
2.32+1.79B\cos\theta$[T]. As shown in Figs.~2(b) and 2(c), we
obtain a remarkably good description of the spike positions $B_c$,
with one adjustable parameter, the Mn concentration $x$ set at
1.37\%, for which $x_{eff} = 1.12$\% and $T_{AF} = 0.47$~K.

%%%%%%%%%%%%%%% figure 2%%%%%%%%%%%%%%%%%%%%%%%%%%%%%%%%%%%%%%%%%%%%
\begin{figure}[t]
\centerline{\epsfig{file=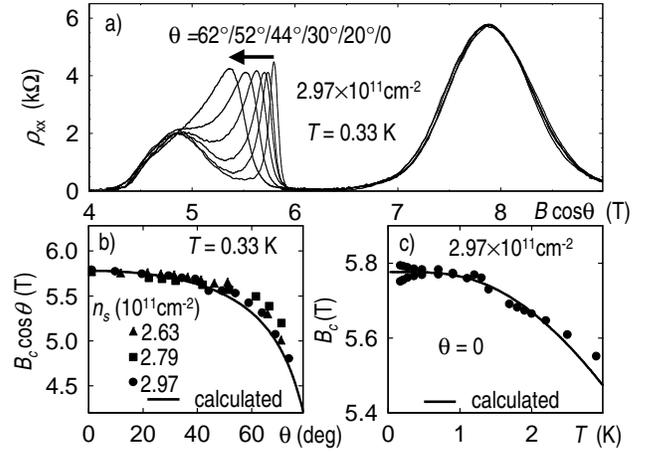,width=8.2cm,clip=}} \caption{(a)
Resistance $\rho_{xx}$ for $n_s=2.97\times10^{11}$~cm$^{-2}$ for
various tilt angles $\theta$ (a). Experimental and calculated
(Eq.~\ref{Brillouin}) spike positions (b,c) for two directions of the
field sweep (c).}
\end{figure}
%%%%%%%%%%%%%%%%%% end figure 2%%%%%%%%%%%%%%%%%%%%%%%%%%%%%%

The interpretation of the resistance spikes in terms of the QHFM
formation implies that they should disappear if $\nu$ deviates from
an integer at $B_c$. This is clearly demonstrated in Fig.~3(a),
where $\rho_{xx}(B)$ is presented for various gate voltages $V_g$,
corresponding to $1.65 \le \nu\le 2.45$ at $B_c = 5.8$~T. A number
of important conclusions emerges from these data. First, in
contrast to the SdH maxima, the spike positions do not depend on
$V_g$, which substantiates the assignment of the spikes to the
level crossing and rules out macroscopic nonuniformities of the
2DES as their origin. Second, the spike amplitude  peaks at $\nu=
2\pm 0.02$, and it decreases for both smaller and higher $\nu$
values, as shown in the inset to Fig.~3. Finally, a careful
examination of the evolution of $\rho_{xx}(B)$ plots with $V_g$
makes it possible to detect spikes at other field values and, by
making use of the fan chart depicted in Fig.~3(b), to identify
indices of the relevant LLs.

Having established that the spikes occur if LLs cross at $\nu$
close to integers, we turn to experimental results which
demonstrate the existence of a phase transition at nonzero
temperatures under such conditions. According to Fig.~4(a), the
spike magnitude exhibits a rather sharp maximum at the temperature
that we identify as $T_C$. At the same time, a hysteresis loop of
$\rho_{xx}(B)$ develops when $B$ is swept in two directions below
$T_C$, as presented in Fig.~4(b). Hence, our results corroborate
the notion \cite{Jung00} that if $\nu$ is close to an integer at
$B_c$, a transition to the Ising QHFM ground state takes place. In
this broken symmetry state, all electrons fill up one LL, leaving
the other LL empty. This is evidenced in Fig.~1(b) which reveals
the absence of a SdH maximum at $B_c$ below $T_C$. However,
depending on a local potential landscape either $0\uparrow$ or
$1\downarrow$ LL is filled up in a given space region. Domain walls
appearing in this way form edge-like channels. Their presence gives
raise to an additional conductance that results in the resistance
spike at $B_c$ \cite{Jung01}.

%%%%%%%%%%%%%%% figure 3 %%%%%%%%%%%%%%%%%%%%%%%%%%%%%%%%%%%%%%%%%%%%
\begin{figure}[tb]
\centerline{\epsfig{file=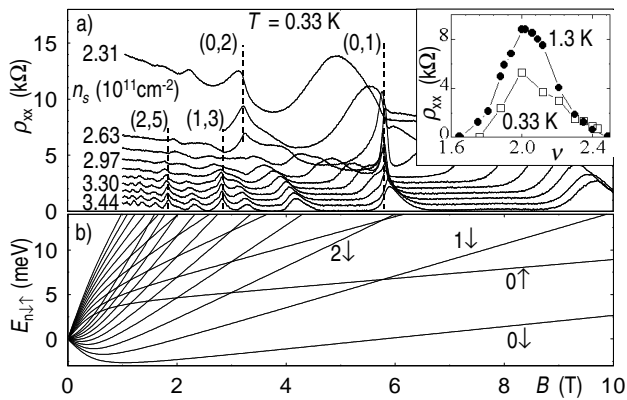,width=8.2cm,clip=}}\caption{(a)
Resistance $\rho_{xx}$ at 0.33~K for  $n_s= 2.31 -
3.44\times10^{11}$~cm$^{-2}$. Traces are shifted vertically for
clarity. Dashed lines mark resistance spikes at the crossing of LLs
with the indices (2,5), (1,3), (0,2), and (0,1) as determined from
LLs energies displayed in (b). The inset shows the height of the
spike at $B_c\simeq5.8$~T as a function of the filling factor.}
\end{figure}
%%%%%%%%%%%%%%%%%% end figure 3%%%%%%%%%%%%%%%%%%%%%%%%%%%%%%%%%%%%%%

The collective nature of the phenomenon observed here points to a
minor importance of the one-electron anticrossing effect driven by
spin-orbit interactions \cite{Falk93} or spin-flip scattering by Mn
ions. Above $T_C$, in turn, the carriers are distributed among two
degenerated subbands, which means that the Fermi level resides in
the vicinity of the DOS maximum at $B_c$. Thus an additional SdH
peak appears at $T>T_C$, as shown in Fig.~1(b), a genuine one
electron effect, examined recently in (Zn,Cd,Mn)Se QWs
\cite{Knob02}.

In order to discuss mechanisms that may control the magnitude of
$T_C$ in our system, we begin with the s$-$d exchange interaction.
According to the mean-field theory of magnetic polarons and Zener's
ferromagnetism \cite{Diet94a,Diet01} we expect the gain of the free
energy per one electron associated with the formation of the QHFM
state to be given by
\begin{equation}\label{Zener}
J_{S} =\frac{\alpha^2}{2k_B(2g\mu_B)^2} \int d\bm{\rho}\int dz
\chi_{\parallel}(z)|\psi(\bm{\rho})\varphi(z)|^4.
\end{equation}
Here, $\chi_{\parallel} =\partial M/\partial H$ is the longitudinal
susceptibility of the Mn spins inside the QW and the integral over
$|\psi(\bm{\rho})\varphi(z)|^4$ is the inverse participation volume
$P$. Recent work on the Knight shift of Mn spin resonance in a
similar (Cd,Mn)Te QW \cite{Tera02} provides information on $P$.
Importantly, the Mn spin resonance was detected resistively, which
means that only those Mn spins that are coupled to the electrons
participating in the charge transport were probed. The data, taken at
low $T$, where motional narrowing ceases to be important, point to a
sevenfold enhancement of $P^{-1}$ for $\nu=3$ over the value expected
for the uniform distribution of the electrons in the QW. In spite of
this enhancement, we find that $J_S$ is rather small, $J_S \approx
0.02$~K at 2~K and at 5.8~T, and decreases even further when $T$
decreases. Another modeling of $\chi_{\parallel}$, for instance by
adding a contribution from the magnetization step due to the nearest
neighbor Mn pairs \cite{Isaa88}, increases $J_S$ by no more than a
factor of two. We note also that the application of $B$, which
saturates the Mn spins, $Sg\mu_BB \gg k_B(T+T_{AF})$, shifts the Mn
excitation spectrum to energies greater than $k_{B}T$. Since in this
regime the Mn spins fluctuate faster than the electronic spins, the
s$-$d coupling can be regarded as mediating interactions among the
itinerant electrons \cite{Sach89}. In such a case, the contribution
of quantum fluctuations to $J_{S}$ becomes important, which further
increases $\chi_{\parallel}$ in Eq.~(\ref{Zener}) in the limit $T
\rightarrow 0$ \cite{Diet94a}. Though the exact value of the
enhancement factor is unknown, we claim nevertheless that the s$-$d
coupling does not give the quantitatively important contribution to
the QHFM stability. Accordingly, our results can be employed directly
to test the existing theoretical predictions on the QHFM stability
\cite{Jung00,Jung01} in the presence of disorder
\cite{Fogl95,Sino00}.

%%%%%%%%%%%%%%% figure 4 %%%%%%%%%%%%%%%%%%%%%%%%%%%%%%%%%%%%%%%%%%%%
\begin{figure}[tb]
\centerline{\epsfig{file=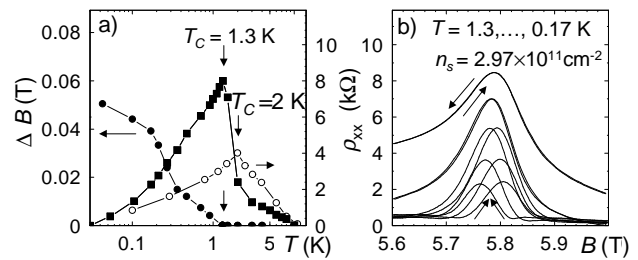,width=8.2cm,clip=}}
\caption{(a) The spike height  as a function of $T$ for $n_s=2.97$
(squares) and $2.54\times10^{11}$~cm$^{-2}$ (open circles).  If
the spike and the SdH peak overlap, the contribution of the latter
is subtracted. Full circles show the width of hysteresis loops
presented in (b), where  $\rho_{xx}(B)$ is depicted  in the region
of the spike for two directions of the field sweep at various $T$.
The sweep rate is 0.3~T/min. The resistance behavior, reminiscent
of critical scattering,  and the presence of magnetic hysteresis
are taken as evidences for the phase transition to QHFM.}
\end{figure}
%%%%%%%%%%%%%%%%%% end figure 4%%%%%%%%%%%%%%%%%%%%%%%%%%%%%%%%%%%%%%

We compute the gain in the energy associated with the formation of
the QHFM state $J=U_{zz}/2$ from the Hartree-Fock theory
\cite{Jung00}, in which effects of remote LLs are neglected. For
the (Cd,Mn)Te dielectric constant $\epsilon=10$ and the envelope
function in the form $\varphi(z) =(2/L_W)^{1/2}\cos(\pi z/L_W)$, we
obtain $J\approx12$~K per electron at $B_c$, whereas $J\approx18$~K
in the limit $\varphi(z) = \delta(0)$. The value of $J$ is
therefore ten times larger than the one-electron anticrossing
energy observed in a similar QW by optical absorption
\cite{Karc01}. Since in the case of the Ising QHFM there are no
skyrmion excitations \cite{Sino00}, the effect of disorder can be
evaluated within the mean-field approach \cite{Fogl95}. In the
presence of a short-range disorder potential, this theory
\cite{Fogl95} predicts the QHFM to occur if $J>W$, where $W=
\hbar[(\pi e\omega_c/2\mu m^*)]^{1/2}/4k_B\approx 8$~K for our QW.
Furthermore, it has recently been suggested \cite{Jung01} that the
magnitude of $T_C$ can actually be much smaller than $J$, as it is
determined by the free energy competition between the magnetic
domain configurational entropy at $T>0$ and the energy penalty
associated with the domain wall formation. Thus, $T_C$ of 2~K that
we observe is consistent with the existing theories of the QHFM
\cite{Jung00,Jung01,Fogl95}. On the other hand, the fact that this
value is four times higher than that of AlAs \cite{Poor00,Jung01}
appears surprising, especially if we note that the crossing of the
LLs $n=0$ and $n=2$ (AlAs) should result in a greater $J$ than in
our case involving $n=0$ and $n=1$ LLs \cite{Jung00}. We assign the
greater stability of the QHFM state in (Cd,Mn)Te to seven times
larger distance to the nearest LLs at $B_c$ (caused by larger
$\omega_c$), which leads to a significantly weaker screening of the
Coulomb interaction compared to AlAs. Furthermore, it is possible
that quenched spin-dependent disorder leads to an increase of the
magnetic stiffness by pinning of the domains in magnetic QWs.
Indeed, we evaluate the corresponding energy scale $W_S$ to be of
the order of 1~K at $B_c$ for the randomly distributed Mn ions.
Within the domain formation scenario \cite{Jung01}, the resistance
spikes correspond to the enhancement of  conductance due to
electron transport along the domain walls, whose total length
diverges at $T\rightarrow T_C^-$ in a narrow field range
corresponding to the demagnetized state \cite{Jung01}. Thus, the
critical behavior of the spike  observed by us provides an
important experimental support for this model of the spike origin.

In conclusion, our results demonstrate that the presence of the
magnetic ions allows one to reach conditions for the observation of
the Ising quantum Hall ferromagnet. Our data, taken over a wide
range of the parameter space, have provided the important
verification of the recent theory of this phase \cite{Jung00} and
for the mechanism of the spike formation \cite{Jung01}. The
arguments have been put forward indicating that the Mn spin
fluctuations give rise to an additional electron-electron
interaction and constitute domain pining centers. It appears at
this point, however, that the Coulomb interaction dominates and
makes the QHFM state more stable than anticipated previously.

This work was supported by the Foundation for Polish Science, KBN
Grants 2-P03B-02417 and 7-T08A-04721, NSF Grant DMR-0071668, and
NHMFL through NSF Cooperative Agreement DMR-0084173. We are
grateful to W.~Plesiewicz, T.~Murphy, and E.~Palm for their
cryogenic expertise.


\begin{thebibliography}{99}
\expandafter\ifx\csname
natexlab\endcsname\relax\def\natexlab#1{#1}\fi
\expandafter\ifx\csname bibnamefont\endcsname\relax
  \def\bibnamefont#1{#1}\fi
\expandafter\ifx\csname bibfnamefont\endcsname\relax
  \def\bibfnamefont#1{#1}\fi
\expandafter\ifx\csname citenamefont\endcsname\relax
  \def\citenamefont#1{#1}\fi
\expandafter\ifx\csname url\endcsname\relax
  \def\url#1{\texttt{#1}}\fi
\expandafter\ifx\csname urlprefix\endcsname\relax\def\urlprefix{URL
}\fi \providecommand{\bibinfo}[2]{#2}
\providecommand{\eprint}[2][]{\url{#2}}


\bibitem[{\citenamefont{DasSarma}(1997)}]{Dass97}
{\it Perspectives
 in Quantum Hall Effects}, edited by S. Das Sarma and A. Pinczuk
(Wiley, New York, 1997).

\bibitem[{\citenamefont{Girvin}(2000)}]{Girv00}
\bibinfo{author}{\bibfnamefont{G.~F.}~\bibnamefont{Giuliani}}
\bibnamefont{and}
  \bibinfo{author}{\bibfnamefont{J.~J.} \bibnamefont{Quinn}},
  \bibinfo{journal}{Phys. Rev. B} \textbf{\bibinfo{volume}{31}},
  \bibinfo{pages}{6228} (\bibinfo{year}{1985});
\bibinfo{author}{\bibfnamefont{S.~M.}~\bibnamefont{Girvin}},
  \bibinfo{journal}{Phys. Today} \textbf{\bibinfo{volume}{53}},
  \bibinfo{pages}{39} (\bibinfo{year}{June 2000}),
\bibinfo{note}{and references therein}.

\bibitem[{\citenamefont{Jungwirth and MacDonald}(2000)}]{Jung00}
\bibinfo{author}{\bibfnamefont{T.}~\bibnamefont{Jungwirth}}
\bibnamefont{and}
  \bibinfo{author}{\bibfnamefont{A.~H.} \bibnamefont{MacDonald}},
  \bibinfo{journal}{Phys. Rev. B} \textbf{\bibinfo{volume}{63}},
  \bibinfo{pages}{035305} (\bibinfo{year}{2000}).

\bibitem[{\citenamefont{Piazza et~al.}(1999)}]{Piaz99}
\bibinfo{author}{\bibfnamefont{S.} \bibnamefont{Koch}},
 \bibinfo{author}{\bibfnamefont{R.~J.} \bibnamefont{Haug}},
 \bibinfo{author}{\bibnamefont{K.~v.} \bibfnamefont{Klitzing}},
 \bibnamefont{and}
 \bibinfo{author}{\bibfnamefont{M.} \bibnamefont{Razeghi}},
  \bibinfo{journal}{Phys. Rev. B}
  \textbf{\bibinfo{volume}{47}}, \bibinfo{pages}{4048}
  (\bibinfo{year}{1993});
  \bibinfo{author}{\bibfnamefont{V.} \bibnamefont{Piazza}}
  \textit{\bibnamefont{et~al.}}, \bibinfo{journal}{Nature}
  \textbf{\bibinfo{volume}{402}}, \bibinfo{pages}{638}
  (\bibinfo{year}{1999});
\bibinfo{author}{\bibfnamefont{J.} \bibnamefont{Eom}}
  \textit{\bibnamefont{et~al.}}, \bibinfo{journal}{Science}
  \textbf{\bibinfo{volume}{289}}, \bibinfo{pages}{2320}
  (\bibinfo{year}{2000});
\bibinfo{author}{\bibfnamefont{J.~H.} \bibnamefont{Smet}}
  \textit{\bibnamefont{et~al.}}, \bibinfo{journal}{Phys. Rev. Lett.}
  \textbf{\bibinfo{volume}{86}}, \bibinfo{pages}{2412}
  (\bibinfo{year}{2001});
  \bibinfo{author}{\bibfnamefont{K.} \bibnamefont{Muraki}},
  \bibinfo{author}{\bibfnamefont{T.} \bibnamefont{Saku}},
  \bibnamefont{and}
 \bibinfo{author}{\bibfnamefont{Y.} \bibnamefont{Hirayama}},
 \bibinfo{journal}{Phys. Rev. Lett.}
  \textbf{\bibinfo{volume}{87}}, \bibinfo{pages}{196801}
  (\bibinfo{year}{2001}).

\bibitem[{\citenamefont{{De Poortere} et~al.}(2000)}]{Poor00}
\bibinfo{author}{\bibfnamefont{E.~P.} \bibnamefont{{De Poortere}}}
  \textit{\bibnamefont{et~al.}}, \bibinfo{journal}{Science}
  \textbf{\bibinfo{volume}{290}}, \bibinfo{pages}{1546}
  (\bibinfo{year}{2000}).

\bibitem[{\citenamefont{Jungwirth and MacDonald}(2001)}]{Jung01}
\bibinfo{author}{\bibfnamefont{T.}~\bibnamefont{Jungwirth}}
 \bibnamefont{and}
  \bibinfo{author}{\bibfnamefont{A.~H.} \bibnamefont{MacDonald}},
  \bibinfo{journal}{Phys. Rev. Lett.} \textbf{\bibinfo{volume}{87}},
  \bibinfo{pages}{216801} (\bibinfo{year}{2001}).

\bibitem[{\citenamefont{Andrearczyk et~al.}(2001)
\citenamefont{T. Andrearczyk,
J. Jaroszy\'nski, M. Sawicki, Le Van Khoi, T. Dietl, D. Ferrand, C.
Bourgognon, J. Cibert, S. Tatarenko, T. Fukumura, Zhengwu Jin, H.
Koinuma, M. Kawasaki}}]{Andr01}
\bibinfo{author}{\bibfnamefont{T.}~\bibnamefont{Andrearczyk}}
\textit{\bibnamefont{et~al.}},
\bibinfo{journal}{{\em Procceedings of the 25th
International Conference on Physics of
Semicondonductors}, Osaka, Japan, 2000, edited by N. Miura and T.
Ando (Springer, Berlin, 2001) p. 235}.

\bibitem[{\citenamefont{Grabecki et~al.}(1993)}]{Grab93}
\bibinfo{author}{\bibfnamefont{G.} \bibnamefont{Grabecki}}
  \textit{\bibnamefont{et~al.}}, \bibinfo{journal}
  {Semicon. Sci. Technol.}
  \textbf{\bibinfo{volume}{8}}, \bibinfo{pages}{S95}
  (\bibinfo{year}{1993}).

\bibitem[{\citenamefont{Smorchkova et~al.}(1997)}]{Smor97}
\bibinfo{author}{\bibfnamefont{I.~P. Smorchkova, N. Samarth,
J. M. Kikkawa, and D. D.
Awschalom}},
   \bibinfo{journal}{Phys. Rev. Lett.}
  \textbf{\bibinfo{volume}{78}}, \bibinfo{pages}{3571}
   (\bibinfo{year}{1997}).

\bibitem[{\citenamefont{Jaroszy\'nski et~al.}(2000)}]{Jaro00}
\bibinfo{author}{\bibfnamefont{J.}~\bibnamefont{Jaroszy\'nski}}
  \textit{\bibnamefont{et~al.}}, \bibinfo{journal}{Physica (Amsterdam) }
  \textbf{\bibinfo{volume}{6E}}, \bibinfo{pages}{790}
   (\bibinfo{year}{2000}).

\bibitem[{\citenamefont{Karczewski et~al.}(1998)}]{Karc98}
\bibinfo{author}{\bibfnamefont{G.}~\bibnamefont{Karczewski}}
  \textit{\bibnamefont{et~al.}}, \bibinfo{journal}{J.\ Cryst. Growth}
  \textbf{\bibinfo{volume}{184/185}}, \bibinfo{pages}{814}
  (\bibinfo{year}{1998}).

\bibitem[{Die()}]{Diet94}
\bibinfo{note}{For a review on DMS, see, {\em e.~g.},
T.\ Dietl, in {\it Handbook on
  Semiconductors}, edited by \ T.S.\ Moss, (North-Holland,
  Amsterdam, 1994) vol.\ 3b,
  p.\ 1251; J. K.\ Furdyna, J.\ Appl.\ Phys. {\bf 64}, R29 (1988)}.

\bibitem[{\citenamefont{Hu et~al.}(1998)}]{Hu98}
\bibinfo{author}{\bibfnamefont{C.~Y.} \bibnamefont{Hu}}
\textit{\bibnamefont{et~al.}},
  \bibinfo{journal}{Phys. Rev. B} \textbf{\bibinfo{volume}{58}},
  \bibinfo{pages}{R1766} (\bibinfo{year}{1998}).

\bibitem[{\citenamefont{Gaj et~al.}(1994)\citenamefont{Gaj,
Grieshaber, Bodin-Deshayes,
Cibert, Feuillet, Merle d'Aubign\'e, Wasiela}}]{Gaj94}
\bibinfo{author}{\bibfnamefont{J.~A.}~\bibnamefont{Gaj}}
\textit{\bibnamefont{et~al.}},
\bibinfo{journal}{Phys. Rev. B} \textbf{\bibinfo{volume}{50}},
\bibinfo{pages}{5512} (\bibinfo{year}{1994}).



\bibitem[{\citenamefont{Fal'ko}(1993)}]{Falk93}
\bibinfo{author}{\bibfnamefont{V.~I.}~\bibnamefont{Fal'ko}},
\bibinfo{journal}{J. Phys.: Cond. Matter}
  \textbf{\bibinfo{volume}{5}}, \bibinfo{pages}{8725}
  (\bibinfo{year}{1993}).


  \bibitem[{\citenamefont{Knobel et~al.}(2002)
  \citenamefont{Knobel, Samarth,
  Harris, and Awschalom}}]{Knob02}
\bibinfo{author}{\bibfnamefont{R.}~\bibnamefont{Knobel}},
  \bibinfo{author}{\bibfnamefont{N.}~\bibnamefont{Samarth}},
  \bibinfo{author}{\bibfnamefont{J.~G.~E.} \bibnamefont{Harris}},
  \bibnamefont{and} \bibinfo{author}{\bibfnamefont{D.~D.}
  \bibnamefont{Awschalom}}, \bibinfo{journal}{Phys. Rev. B}
  \textbf{\bibinfo{volume}{65}}, \bibinfo{pages}{235327}
  (\bibinfo{year}{2002}).

\bibitem[{Dietl()}]{Diet94a}
\bibinfo{note}{See, T.\ Dietl, in Ref.~12, pp.~1287-96}.

\bibitem[{\citenamefont{Dietl et~al.}(2001)\citenamefont{Dietl and
  Ohno}}]{Diet01}
\bibinfo{author}{\bibfnamefont{T.}~\bibnamefont{Dietl}}
\bibnamefont{and} \bibinfo{author}{\bibfnamefont{H.}~\bibnamefont{Ohno}},
\bibinfo{journal}{Physica (Amsterdam) } \textbf{\bibinfo{volume}{9E}},
  \bibinfo{pages}{185} (\bibinfo{year}{2001}).

\bibitem[{\citenamefont{Teran et~al.}(2002)\citenamefont{F. J. Teran,
T. Andrearczyk,
J. Jaroszy\'nski, G. Karczewski, T. Wojtowicz, T. Dietl, E. Papis, E.
Kami\'nska, A. Piotrowska}}]{Tera02}
\bibinfo{author}{\bibfnamefont{F.~J.}~\bibnamefont{Teran}}
\textit{\bibnamefont{et~al.}},
\bibinfo{journal}{Physica (Amsterdam) } \textbf{\bibinfo{volume}{12E}},
\bibinfo{pages}{356} (\bibinfo{year}{2002}).

\bibitem[{\citenamefont{Isaacs et~al.}(1984)\citenamefont{Isaacs}}]{Isaa88}
\bibinfo{author}{\bibfnamefont{E.~D.}~\bibnamefont{Isaacs}},
\textit{\bibnamefont{et~al.}},
  \bibinfo{journal}{Phys. Rev. B}
  \textbf{\bibinfo{volume}{38}}, \bibinfo{pages}{8412}
   (\bibinfo{year}{1988}).

\bibitem[{\citenamefont{Sachdev}(1989)\citenamefont{Sachdev}}]{Sach89}
\bibinfo{author}{\bibfnamefont{S.}~\bibnamefont{Sachdev}},
\bibinfo{journal}{Phys. Rev. B} \textbf{\bibinfo{volume}{39}},
\bibinfo{pages}{5297} (\bibinfo{year}{1989}).

\bibitem[{\citenamefont{Fogler et~al.}(1995)
\citenamefont{Fogler, Shklovskii}}]{Fogl95}
\bibinfo{author}{\bibfnamefont{M.~M.} \bibnamefont{Fogler}}
 \bibnamefont{and} \bibinfo{author}{\bibfnamefont{B.~I.}
 \bibnamefont{Shklovskii}},
\bibinfo{journal}{Phys. Rev. B} \textbf{\bibinfo{volume}{52}},
\bibinfo{pages}{17~366} (\bibinfo{year}{1995}).

\bibitem[{\citenamefont{Sinova et~al.}(2000)
\citenamefont{Sinova,MacDonald,Girvin}}]{Sino00}
\bibinfo{author}{\bibfnamefont{J.} \bibnamefont{Sinova}},
  \bibinfo{author}{\bibfnamefont{A.~H.} \bibnamefont{MacDonald}},
\bibnamefont{and} \bibinfo{author}{\bibfnamefont{S.~M.}
  \bibnamefont{Girvin}}, \bibinfo{journal}{Phys. Rev. B}
  \textbf{\bibinfo{volume}{62}},
\bibinfo{pages}{13~579} (\bibinfo{year}{2000});
\bibinfo{author}{\bibfnamefont{S.}~\bibnamefont{Rapsch}},
  \bibinfo{author}{\bibfnamefont{J.~T.} \bibnamefont{Chalker}},
  \bibnamefont{and} \bibinfo{author}{\bibfnamefont{D.~K.~K.}
  \bibnamefont{Lee}}, \bibinfo{journal}{Phys. Rev. Lett.}
  \textbf{\bibinfo{volume}{88}}, \bibinfo{pages}{036801}
  (\bibinfo{year}{2002}).

\bibitem[{\citenamefont{Karczewski}(2001)
\citenamefont{Karczewski Y.-J. Wang}}]{Karc01}
\bibinfo{author}{\bibfnamefont{G.}~\bibnamefont{Karczewski}}
\bibnamefont{and}
 \bibinfo{author}{\bibfnamefont{Y.-J.}~\bibnamefont{Wang}},
 \bibinfo{journal}{Physica (Amsterdam) }
 \textbf{\bibinfo{volume}{298B}}, \bibinfo{pages}{402}
 (\bibinfo{year}{2001}).

\end{thebibliography}
\end{document}